\documentclass[12pt]{article}
\usepackage{axodraw}
\def\header{\begin{flushleft} 
            ZU-TH \\November 1999
            \end{flushleft}}

\columnwidth\textwidth

\def\GSW_sign{}

\newcommand{\ba}{\begin{array}}
\newcommand{\ea}{\end{array}}
\newcommand{\bd}{\begin{displaymath}}
\newcommand{\ed}{\end{displaymath}}
\newcommand{\be}{\begin{equation}}
\newcommand{\ee}{\end{equation}}
\newcommand{\bea}{\begin{eqnarray}}
\newcommand{\eea}{\end{eqnarray}}

\def\qb{\bar{q}}
\def\ub{\bar{u}}
\def\db{\bar{d}}

\def\bra{\langle}
\def\ket{\rangle}

\def\a{\alpha}
\def\b{\beta}
\def\g{\gamma}
\def\d{\delta}
\def\e{\epsilon}
\def\ve{\varepsilon}
\def\l{\lambda}
\def\m{\mu}
\def\n{\nu}

\def\r{\rho}
\def\G{\Gamma}
\def\D{\Delta}
\def\L{\Lambda}

\def\t{\tau}
\def\T{\Theta}

\def\to{\rightarrow}

\begin{document}
\thispagestyle{empty}
%\addtocounter{page}{-1}
\header \vspace*{2cm} \centerline{\Large\bf 
Radiative corrections of $O(\a)$ for pion beta decay}
 \centerline{\Large\bf  in the light-front quark model } 
     \vspace*{2.0cm}
\centerline{\large Wolfgang Jaus}
\vspace*{0.5cm} \centerline{Institut 
f\"ur Theoretische Physik der Universit\"at Z\"urich, Winterthurerstr. 190, CH-8057 Z\"urich} 
\centerline{Switzerland}
\vspace{2cm}

%%%%%%%%%%%%%%%%%%% ABSTRACT %%%%%%%%%%%%%%%%%%%%%%%%%%%%%%%%%%%%
\centerline{\Large\bf Abstract} \vspace*{1cm} 
\hspace{0.5cm}If the CKM matrix element $V_{ud}$, that can be derived from
superallowed nuclear decays, neutron decay and pion beta decay, is used for 
a precision test of the unitarity of the CKM matrix, the combination of the
present world data seems to indicate a small violation of the unitarity condition for the first row.
While an accurate calculation of the radiative corrections of $O(\a)$ (RC) is
crucial in order to determine the value of $V_{ud}$ as precisely as possible, the theoretical
analysis has been limited in the past by the rather crude estimate of the effect of
the hadronic structure. Only the contribution due to the axial current depends on the
hadronic environment. We develop a strategy to deal with the influence of the hadronic
structure on the decay properties of the simplest hadron, the pion, and calculate the 
contribution of the axial vector current
to the RC, using a light-front model for the pion. 
Its $q\bar{q}$ bound state structure is well described by two parameters,
constituent quark mass and confinement scale, that have been fixed by a comparison with the data. We take into
consideration three different groups of two-loop diagrams, and derive their light-front
representations. We discuss the associated zero-mode problem and show that the respective light-front 
amplitudes are free of spurious contributions. There is only a small model dependent uncertainty 
of the final result for the RC for pion beta decay.

\newpage
%%%%%%%%%%%%%%%%%%%%%% SECTION 1 %%%%%%%%%%%%%%%%%%%%%%%%%%%%%%%%
\section{Introduction} 
\hspace{0.5cm} For three fermion generations, unitarity of the
Cabibbo-Kobayashi-Maskawa (CKM) matrix requires the sum of the squared
moduli of the first three elements to be equal to one:
\be
V^2 \equiv |V_{ud}|^2 + |V_{us}|^2 + |V_{ub}|^2 = 1.
\ee
A test of this property is of crucial importance since a violation of unitarity
would be evidence for new physics, and the use of such a result to constrain
possible extensions of the Standard Model would require a precise value of $V^2$ and its uncertainty.

The unitarity sum $V^2$
critically depends upon the precise value of the matrix element $V_{ud}$
for the decays $u \to d \bar{e} \nu$ and $d \to u e \bar{\nu}$. These
quark level transitions give rise to superallowed Fermi beta decays, the decay
of the free neutron $n \to p e \bar{\nu}$ and
pion beta decay $\pi^+ \to \pi^0 \bar{e} \nu$. In each case the measured rate
can be used to determine the value of $V_{ud}$, after radiative corrections 
(RC) and the effect of the hadronic environment  have been separated out.

A general formula for the RC of order $ \a$ to the transition rates has
been given by Sirlin \cite{sirlin1}. The total decay rate $1/\t$ can be
separated into the uncorrected expression, denoted by $1/\t_0$, and an an overall factor
as 
\bea
1/\t &=& 1/\t_0 \Big (1+\d) \nonumber \\
\d &=&  \frac{\a}{2\pi} \bigg[g(E_0)+3\ell n\frac{M_Z}{M_p}+A_g
       \bigg] \nonumber \\
&& + \frac{\a}{2\pi} \bigg[ 3(Q_u +Q_d)\ell n\frac{M_Z}{M_A} +2C \bigg],
\eea
where $Q_u$ and $Q_d$ are the quark charges of $u$ and $d$ quarks.
The Sirlin function $g(E,E_0)$ has been defined in \cite{sirlin2}
as a function of the electron or positron energy $E$ and  
represents the RC to the electron or positron spectrum in allowed
beta decay. In the total
decay rate $1/\t$ it is replaced by the averaged value $g(E_0)$;
$E_0$ is the end-point energy of the spectrum.

The correction terms of $O(\a)$ consist of three distinct parts.
The first two terms $g(E_0)+3\ell n(M_Z/M_p)$ represent the contribution of the
vector current and are independent of hadron dynamics. The $Z$ boson mass
$M_Z$ is a consequence of short-distance effects while the proton mass
$M_p$ cancels in the sum of the two terms. The third term $A_g$ is a small
asymptotic QCD correction term;  $A_g = -0.34$ \cite{sirlin1,masi}. Finally there
is a contribution $\ell n(M_Z/M_A)+2C$ induced by the axial vector current,
where the logarithm is again the result of short-distance effects, with 
$M_A$ acting as an effective low-energy cutoff (presumably roughly equal to
the $a_1$ meson mass), and $2C$ stands for the remaining low-energy part.

The value of $M_A$ is uncertain; Marciano and Sirlin \cite{masi} suggested
a range $ 400  MeV \le M_A \le 1600 MeV$, while Sirlin \cite{sirlin4}
proposed an even wider range
\be
M_{a1}/2 \le M_A \le 2M_{a1},
\ee
with the central value at the $a_1$ meson mass $M_{a1} = 1.26  GeV$. The
quantity $2C$ is model dependent and has been calculated in the Born approximation
in Refs. \cite{masi,tow} using nucleon electromagnetic and axial form factors.
For pion beta decay $C=0$ in the Born approximation since the axial vector current
does not couple to a pseudoscalar meson. The resulting values for $C$ are
\bea
C=C_{Born} = 
\left\{ \begin{array}{ll}
0.885 &({\rm  superallowed \quad and \quad neutron \quad decays})\\
0 &({\rm  pion \quad beta \quad decay}). \end{array} \right. 
\eea 
For superallowed beta decays there are additional nuclear structure dependent 
contributions to $C$ which have been proposed and discussed in Ref. \cite{jaus1}.

The uncorrected decay rate $1/\t_0$, defined by Eq.(1.2), still incorporates Coulomb
corrections and $Z$-dependent radiative corrections of $O(Z\a^2)$ and $O(Z^2 \a^3)$
for superallowed nuclear decays, and depends on hadronic form factors, which encode
the effect of the quark structure of the decaying hadron. 

Recently the current status of $V_{ud}$ has been reviewed by Towner and Hardy \cite{toha},
based upon the current world data for the three decay modes indicated above. To date,
nine superallowed $0^+ \to 0^+$ transitions have been measured to $\pm 0.1\%$ precision
or better, and the result for $V_{ud}$ obtained from the average ft value is

\be
|V_{ud}| = 0.9740 \pm 0.0005.
\ee
From this value of $V_{ud}$ the unitarity sum, Eq.(1.1), becomes

\be
V^2 = 0.9968 \pm 0.0014,
\ee
where the PDG 98 \cite{PDG98} recommendations for $V_{us}$ and $V_{ub}$ have 
been used in Ref.\cite{toha}.
The value for the CKM matrix element $V_{us}$ determined from an analysis of kaon and hyperon
decays is $|V_{us}| = 0.2196 \pm 0.0023$, while the value for $V_{ub}$ is
$|V_{ub}| = 0.0032 \pm 0.0008$ and does not affect the unitarity sum at its present level
of accuracy.

According to the analysis of Towner and Hardy, the error bar associated with the value of
$V_{ud}$ is caused mainly by the uncertainty in the RC $(\pm 0.0004)$ due to the prescription
(1.3) for the effective low-energy cutoff and the uncertainty in the nuclear isospin
symmetry-breaking correction $(\pm 0.0003)$, while the average experimental uncertainty is quite
small $(\pm 0.0001)$.

The problems associated with a precise treatment of nuclear structure effects can be avoided
if the beta decay of free hadrons is considered instead. A survey of world data on neutron
decay observables has been presented in Ref. \cite{toha} and it has been noted that the
derivation of the value of $V_{ud}$ from $n$ decay is limited largely by the uncertainty in
the overall average value of $\l = g_A /g_V$. However, there is a new result for the beta
asymmetry obtained by the PERKEO II collaboration \cite{perkeo} which leads to the
value $|\l| = 1.2735 \pm 0.0021$. This single value, combined with the world average for 
the neutron lifetime, leads to the following value for $V_{ud}$ \cite{toha}:

\be
|V_{ud}| = 0.9714 \pm 0.0015.
\ee
The unitarity sum is then
\be
V^2 = 0.9919 \pm 0.0030 .
\ee
The error given in (1.7) is three times larger than the error in (1.5) and is dominated by
the uncertainty in the measurement of the beta asymmetry  but, as in the analysis of the
superallowed decays, still contains the contribution of the uncertainty in the RC.

The results for $V_{ud}$ and the unitarity sum $V^2$ given in Eqs.(1.5)-(1.8)
are consistent with each other and seem to indicate a substantial violation of the unitarity
condition (1.1) for three generations. Moreover, they support the conclusion reached in Ref.
\cite{toha}, that the treatment of
the effect of the nuclear environment in superallowed nuclear decays is reliable, with only 
a small error, and
that there is no evidence that the unitarity problem can be solved by improvements in the
calculation of nuclear structure effects.

In order to obtain more information on the unitarity problem accurate measurements of the 
pion beta decay observables would be of great importance. Like the superallowed nuclear decays
pion beta decay is a pure vector transition and the matrix element of the axial vector current,
which complicates the analysis of neutron decay, does not contribute to the lowest order
amplitude. In higher orders both the vector and the axial vector parts of the weak current
contribute. The expression for the radiative corrections of $O(\a)$ is given in Eq.(1.2).
Moreover, since the decaying pion is free, the nuclear structure dependent corrections that
complicate nuclear beta decay are absent. Based on the lifetime \cite{PDG98}

\be
\t_{exp} =(2.6033 \pm 0.0005) \times 10^{-8} s
\ee
and the branching ratio \cite{mc far}

\be
BR = (1.025 \pm 0.034) \times 10^{-5},
\ee
the value of $V_{ud}$ was determined in Ref. \cite{toha} to be

\be
|V_{ud}| =0.9670 \pm 0.0161
\ee
and the unitarity sum

\be
V^2 = 0.9833 \pm 0.0311.
\ee
The price to pay for the advantage of a simple theoretical analysis
of pion beta decay is a large error in $V_{ud}$ due to the considerable
experimental difficulty in  measuring the $\pi$ branching ratio with a
precision comparable to the one obtained in superallowed beta decays.
However, there is a proposal for an experiment at PSI \cite{pibeta}
with the aim of making a precise determination of the pion beta decay
rate. In the first phase of the experiment it is intended to measure the
branching ratio with an accuracy of $0.5 \%$. The proposed experimental method
was designed to finally achieve an overall level of uncertainty in the range
of $0.2 - 0.3 \%$.

The decay rate for pion beta decay including the RC of order $\a$ is
given by Eq. (1.2), where an approximate expression for
the uncorrected decay rate has been derived long ago by K$\ddot{a}$ll$\acute{e}$n \cite{kallen}:

\bea
1/\t_0&=& \frac{G^2_F |V_{ud}|^2}{30 \pi^3} \bigg (1- \frac{\D}{2M_+} \bigg )
        \D^5 f(\e,\D), \\
f(\e,\D) &=& \sqrt{1-\e} \, \bigg [ 1-\frac{9\e}{2}-4\e^2 \nonumber \\
&+& \frac{15}{2} \e^2 \ell n \Big ( \frac{1+\sqrt{1-\e}}{\sqrt{\e}} \Big )
-\frac{3}{7} \frac{\D^2}{(M_+ +M_0)^2} \bigg ],
\eea
with $\e = m^2_e/\D^2$ and $\D = M_+ - M_0$, where $M_+$ and $M_0$ are the masses of $\pi^+$ and
$\pi^0$; $G_F$ is the Fermi coupling constant.
Equation (1.14) includes the leading correction in an expansion in powers of
$\D^2/(M_+ +M_0)^2$ \cite{sirlin1}. The effect of the quark structure has been
neglected entirely, and in Sect.2 we shall study the error made by this approximation.
In particular, we shall investigate the effect of isospin violation due to the
quark mass difference $m_d -m_u$, in order to make sure that isospin breaking effects
do not produce unexpectedly large contributions.

For a precision test of the unitarity of the CKM matrix, i.e. of the Standard 
Model, an accurate calculation of the RC, in particular a reliable determination
of the effect of the hadronic structure, is crucial. 
The terms in the electromagnetic radiative corrections of $O(\a)$ that are
generated by the vector current (the first two terms in (1.2)) are firmly founded on a
current algebra formulation and the details of the underlying quark structure are of only
minor importance. We shall not further consider that part of the RC of $O(\a)$ which
is induced by the vector current.
While the short-distance contribution of the axial vector current is well
established too, its role at low energies strongly depends upon the detailed quark structure
of the decaying hadron and its influence on the decay properties has been estimated
only very roughly in terms of an effective low-energy cutoff $M_A$ and the
quantity $C$. We do not know of any published work that attempts to obtain
the contribution of the axial vector current using a model of hadronic structure.  
However, it is evident that a reliable interpretation of the experimental data and a
conclusive
analysis of the unitarity problem necessarily requires a more refined treatment of the effect
of the quark structure in order to substantially reduce the theoretical uncertainties and
to firmly establish the size of the hadronic corrections.

In this paper we shall calculate the axial vector contribution to the RC in the case of
pion beta decay in the framework of the light-front quark model (LFQM), which is
a relativistic constituent quark model based on the light-front
formalism \cite{terentev}. The LFQM provides a conceptually simple, 
phenomenological method for the determination of hadronic form factors
and coupling constants, and has become a much used tool for investigating
various electroweak properties of light and heavy mesons (see e.g.
\cite{jaus2,choi} and references therein). In Ref.\cite{jaus3} we have
presented a covariant extension of the LFQM which permits the calculation
of all the form factors that are necessary to represent the Lorentz
structure of a hadronic matrix element. In this approach a meson is
composed of valence quarks with constituent quark masses and the structure
of the bound $q\qb$ meson
state is approximated by a covariant model vertex function, which depends
on a parameter $1/\b$ which essentially determines the confinement scale,
i.e. the size of the composite meson. Form factors are given in the one-loop
approximation as light-front momentum integrals. As an example, it was shown in 
Ref. \cite{jaus3} that a prediction of the electromagnetic form factor of the pion
for small values of the momentum transfer can be made that is in
good agreement with the data.

The simple structure of the $q\qb$ bound state should allow definite
conclusions about the relative importance of the hadronic environment
in a calculation of the RC. Radiative corrections of order $\a$ to the form
factors that describe pion beta decay arise from the virtual exchange
of $Z$, $\g$ or $W$ and are represented by two-loop diagrams. We shall
extend the approach of Ref.\cite{jaus3} and derive unique LFQM expressions
(that are free of spurious contributions) for the two-loop
diagrams associated with the axial vector current, and derive in this way
the effect of the hadronic structure on the $O(\a)$ corrections
for pion beta decay. This determination of the effect of the hadronic
environment by means of a two-loop calculation should be just as reliable
as the one-loop calculation of the electromagnetic form factor of the
pion.

We shall show in this work that the uncertainty of the hadronic corrections due to the particular quark
structure of the pion is small for pion beta decay. This result is in contrast
to the situation for superallowed nuclear decays and neutron decay where the large
value of $C$, Eq.(1.4), signals a much greater importance of the detailed quark structure
with all its model dependent uncertainty. We shall analyze superallowed nuclear decays
and neutron decay in a future work in a similar manner as for pion beta decay. But even
without knowing the result of such an investigation it is clear that pion beta decay,
once precise data are available, will always have a unique position due to the simple
quark structure of the pion which generates hadronic corrections with very small
uncertainties.

In Sect.2 we present the general formalism for pion beta decay without radiative corrections,
which is analyzed in terms of two form factors that describe the quark structure of the pion.
We investigate the effect on the decay rate of both the isospin violation due to the quark mass difference and
the momentum transfer dependence of the form factors. In Sect.3 the detailed calculation of
the RC due to the axial vector current is presented. We consider three different groups of two-loop
diagrams, and derive their light-front representations. We discuss the associated zero-mode
problem in the Appendix and show that the respective light-front amplitudes are unique, i.e.
free of spurious contributions. We approximate higher order gluon exchange effects by means
of $\r$ exchange diagrams, which are shown to be of only minor importance if appropriate off-shell
form factors are used. Sect.4 contains our result for the RC for pion beta decay.

%%%%%%%%%%%%%%%%%%%% SECTION 2 %%%%%%%%%%%%%%%%%%%%%%%%%%%%%%%%
\section{General formalism for pion beta decay without radiative corrections}
\setcounter{equation}{0}

\hspace{0.5cm} The amplitude without radiative corrections for the decay 
$\pi^+ \to \pi^0 \bar{e} \nu$ is given by

\be
T_1 = \frac{G_F}{\sqrt{2}} V_{ud}  \bra \, \pi^0 (P'') |\db \g_{\mu}(1-\g_5)u | \pi^+
(P') \, \ket L^{\mu}
\ee
where the matrix element of the leptonic current is

\be
L_{\mu} = \ub_{\nu}(k_{\nu}) \g_{\mu}(1-\g_5) v_e(l)
\ee
and $k_{\nu},l$ are the 4-momenta of the neutrino and the positron respectively. We represent
the hadronic matrix element for pion beta decay in terms of appropriate form factors

\be
\bra \, \pi^0 (P'') |\db \g_{\mu}u | \pi^+ (P') \, \ket = \sqrt{2} \Big\{ (P'+P'')_{\mu} F_1(q^2) +
q_{\mu} F_2(q^2) \Big \}
\ee

\be
\bra \, \pi^0 (P'') |\db \g_{\mu}\g_5 u | \pi^+ (P') \, \ket =0
\ee
where $q=P'-P''$ is the 4-momentum transfer which varies within the range
$m^2_e \le q^2 \le (M_+ - M_0)^2$.

It is convenient to analyze semileptonic decays of pseudoscalar mesons in terms of the form
factors $F_1(q^2)$ and $F_0(q^2)$, where the scalar form factor $F_0(q^2)$ is defined by

\be
F_0 (q^2) = F_1 (q^2) + \frac{q^2}{M^2_+ -M^2_0} F_2 (q^2).
\ee

The differential partial width in terms of these form factors is then
\be
\frac{d \G_0(\pi^+ \to \pi^0 \bar{e} \nu)}{dq^2} = \frac{G^2_F |V_{ud}|^2 M^3_+}{32 \pi^3}
\r (q^2)
\ee
and the density $\r (q^2)$ consists of spin 0 and spin 1 contributions as follows
\be
\r (q^2) = \r_0(q^2)+\r_1(q^2)
\ee
\be
\r_0(q^2)= \frac{m^2_e}{q^2} \Big (F_0(q^2) \Big )^2 \Big (1- \frac{m^2_e}{q^2} \Big )^2
\Big (1-\frac{M^2_0}{M^2_+} \Big )^2 \frac{p_{\pi}}{M_+}
\ee

\be
\r_1(q^2)= \frac{8}{3} \Big (F_1(q^2) \Big )^2 \Big (1- \frac{m^2_e}{q^2} \Big )^2
\Big (1+\frac{m^2_e}{2q^2} \Big ) \Big ( \frac{p_{\pi}}{M_+} \Big )^3
\ee
where $p_\pi$ is the recoil momentum of the $\pi^0$ in the $\pi^+$ rest frame:
\be
p^2_{\pi} = \frac{1}{4 M^2_+} \Big \{ (M^2_+ - M^2_0)^2 + q^4 - 2q^2(M^2_+ + M^2_0) \Big \}.
\ee
If the quark structure of the pion is neglected, i.e.
in the limit $F_1(q^2)=F_0(q^2)=1$ and for the approximation $p^2_{\pi}  \simeq
(M_+ +M_0)^2(\D^2 -q^2)/4M^2_+$,
the integrated partial
width leads to the approximate expression for the total decay rate $1/\t_0$, Eq.(1.13).

In this Section we shall briefly discuss the exact integrated partial width (2.6) based upon 
the formulas for the form factors $F_1(q^2)$ and $F_2(q^2)$, which we have derived in the
framework of the quark model in Ref. \cite{jaus3}.

%*********************************************************************************************
%FIGURE 1a,b

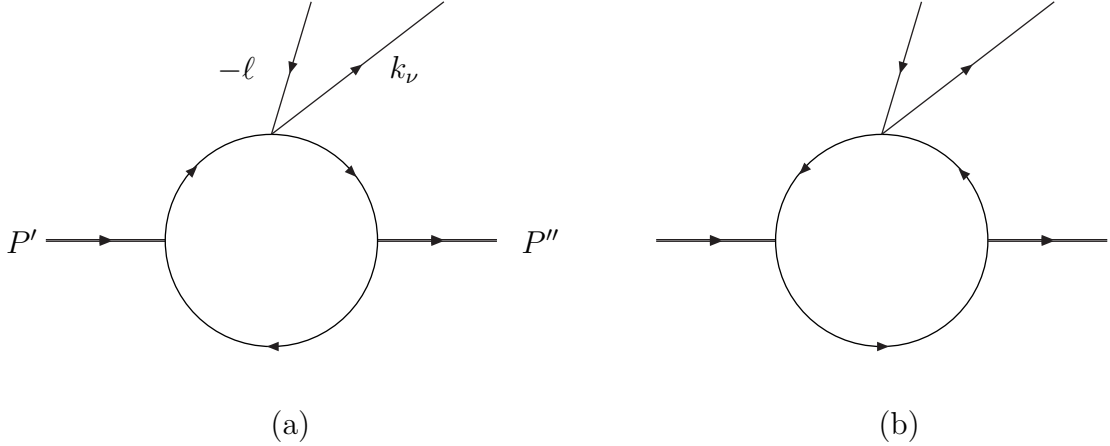
\begin{figure}[h] 
\begin{center}
\begin{picture}(400,180)

%\Line(0,0)(200,0)
%\Line(0,0)(0,180)
%\Line(0,180)(200,180)
%\Line(200,0)(200,180)

%FIGURE 1a

\Text(0,90)[l]{$P'$}
\ArrowLine(15,90.3)(60,90.3)
\ArrowLine(15,89.7)(60,89.7)
\ArrowLine(140,90.3)(185,90.3)
\ArrowLine(140,89.7)(185,89.7)
\Text(195,90)[l]{$P''$}

\GCirc(100,90){40}{1}

\ArrowLine (101,50)(100,50)
\ArrowLine (70,116)(71,117)
\ArrowLine (130,116)(131,115)

\ArrowLine (115,180)(100,130)
\ArrowLine (100,130)(165,180)

\Text(80,155)[l]{$-\ell$}
\Text(145,155)[l]{$k_\n$}

\Text(100,20)[l]{(a)}

%FIGURE 1b

\ArrowLine(245,90.3)(290,90.3)
\ArrowLine(245,89.7)(290,89.7)
\GCirc(330,90){40}{1}
\ArrowLine(370,90.3)(415,90.3)
\ArrowLine(370,89.7)(415,89.7)

\ArrowLine (330,50)(331,50)
\ArrowLine (301,117)(300,116)
\ArrowLine (361,115)(360,116)

\ArrowLine (345,180)(330,130)
\ArrowLine (330,130)(395,180)

\Text(330,20)[l]{(b)}

\end{picture}
\caption{The one-loop contributions to pion beta decay.}
\end{center}
\end{figure}

%*********************************************************************************************

The hadronic matrix element (2.3) is given in the one-loop approximation, corresponding
to the diagrams of Fig.1, as a light-front momentum integral, denoted by $A_{\mu}$. 
The 4-momentum of a meson of mass $M'$ in terms
of light-front components is $P' = (P^{\prime -},P^{\prime +},P'_{\perp})$, where the transverse vector
is $P'_{\perp} = (P^{\prime 1},P^{\prime 2} )$. Its constituent quarks have masses $m'_1$, $m_2$ and
4-momenta $p'_1$, $p_2$, respectively, and the total 4-momentum of the meson state is
given by $p'_1 +p_2 =P'$, i.e. the quarks are in general off the mass-shell. 
The appropriate variables for
the internal motion of the constituents, $(x,p'_\perp)$, are defined by
\bea
p'^+_1 &=& xP'^+ \qquad , \quad p_2 ^+ = (1-x)P'^+ \nonumber\\
p'_{1 \perp} &=& xP'_\perp + p'_\perp \quad , \quad p_{2 \perp} = (1-x)P'_\perp - p'_\perp \nonumber
\eea
and the kinematic invariant mass is
\be
M'^2_0 =  \frac{p'^2_\perp  + m'^2_1  }{x}
                         + \frac{p'^2_\perp  + m_2 ^2 }{1-x} .
\ee

For the transition between an initial $\pi^+ = u\db$ with 4-momentum $P'$, mass $M'$, and
internal variables and masses of its constituent quarks $(x,p'_\perp ,m'_1 ,m_2 )$ and a final 
$\pi^0 =(d\db -u\ub)/\sqrt{2}$ with 4-momentum $P''$, mass $M''$, 
and the corresponding internal quantities 
$(x,p''_\perp ,m''_1 ,m_2 )$, the momentum
integral $A_\m$, in a Lorentz frame with $q^+ =0$,
consists of two parts that describe  the $u \to d$ transition of Fig.1a and the
$\db \to \ub$ transition of Fig.1b, and is given by
\be
A_\m = \frac{1}{\sqrt{2}} \Big ( H_{\mu}(m_u,m_d,m_d)+H_{\mu}(m_d,m_u,m_u) \Big )
\ee
where
\be
H_\m(m'_1,m''_1,m_2) = \frac{N_c}{16\pi^3} \int^1_0 dx \int d^2 p'_\perp \frac{h'_0h''_0}
{(1-x)N'_1 N''_1}S_\m
\ee
with
\be
S_\m = tr\left[\g_5(\not p''_1+m''_1)\g_\m(\not p'_1+m'_1)\g_5(-\not p_2+m_2)
\right],
\ee
where $N_c$ is the number of colors, i. e. $N_c =3$.
The light-front momentum integral $H_\m$, Eq.(2.13), is computed at the pole of the spectator quark:
\be
N_2 \equiv p^2_2 -m^2_2 =0.
\ee
In our formalism \cite{jaus3} four-momentum is conserved and the 4-vectors appearing in the
trace (2.14) are then given by
\bea
p_2 &=&(\frac{m^2_{2\perp}}{p^+_2},p^+_2,p_{2\perp}) \nonumber\\
p'_1 &=&P'-p_2 \\
p''_1 &=& p'_1 -q, \nonumber
\eea
where $m^2_{2 \perp} = m^2_2 + p'^2_\perp$. It follows from Eq.(2.16) that
\bea
N'_1 & \equiv &p'^2_1 -m'^2_1 =x(M'^2-M'^2_0 ) \nonumber \\
N''_1& \equiv &p''^2_1-m''^2_1=x(M''^2-M''^2_0) \\
M''^2_0&=&\frac{p''^2_\perp+(1-x)m''^2_1+xm^2_2}{x(1-x)}, \nonumber
\eea
and $p''_\perp = p'_\perp -(1-x)q_\perp$. In our phenomenological approach we have
chosen a pseudoscalar vertex operator for the $q\qb$ pair bound in a S-state state, 
with the matrix structure
of $\g_5$ and vertex functions $h'_0$ and $h''_0$, where \cite{jaus3}
\be
h'_0 =h'_0(M'_0)= \left[ \frac{M'^4_0 -(m'^2_1 -m^2_2 )^2}{4 M'^3_0} \right]^{1/2}
\frac{M'^2-M'^2_0}{[M'^2_0 -(m'_1 -m_2 )^2 ]^{1/2}} \phi (M'^2_0 ) \\
\ee
for the $q\bar{q}$ bound state of mass $M'$,
and a similar equation for $h''_0$. The orbital wave function is assumed
to be a simple function of the kinematic invariant mass as
\be
\phi (M'^2_0 ) = N' exp(-M'^2_0 /8 \b'^2 ),
\ee
where $N'$ is the normalization constant and the parameter $1/\b'$
determines the confinement scale.
The normalization condition is obtained for $M' =M''$, $\b ' =\b ''=\b$, $m'_1=m''_1=m_2=m$ 
and $q^2=0$, either as a relation for $H_\m(m,m,m)$:
\be
H_\m (m,m,m) = (P'+P'')_\m,
\ee
or as a relation for the orbital wave function:
\be
\frac{N_c}{16\pi^3} \int^1_0 dx \int d^2 p'_\perp \frac{M'_0}{2x(1-x)}
| \phi(M'^2_0)|^2 = 1,
\ee
which for the equal mass case is given explicitly by
\be
\phi(M'^2_0)=\pi^{-3/4}\b^{-3/2} \Big (\frac{8\pi^3}{3} \Big )^{1/2}
exp \Big( -(M'^2_0 -4m^2)/8\b^2 \Big ). \nonumber
\ee

While the form factor $F_1(q^2)$ in the one-loop approximation can be derived directly from
the plus component of the momentum integral $A_\m$ (2.12), the calculation of the form factor
$F_2(q^2)$ requires an appropriate account of the effect of zero-modes, as we have shown
in \cite{jaus3}. We shall not write down the formulas for the form factors, they can be found
in Ref.\cite{jaus3}, but quote the results of the numerical calculation. In the limit of 
exact isospin symmetry the quark masses and the pion masses are equal, i.e. $m_u=m_d=m$ and 
$M_+ =M_0 =M_\pi$, and the form factors can be predicted:
$F_1(q^2)=F_{\pi}(q^2)$, where $F_{\pi}$ is the charge form factor of the pion with
$F_{\pi}(0)=1$, and $F_2(q^2)=0$. In our model the effect of isospin symmetry breaking is
generated by a finite quark mass difference $\D m =m_d -m_u$, while the parameters for the 
wave functions of $\pi^+$ and $\pi^0$ are kept equal: $\b_+ =\b_0 =\b_{\pi}$. We use the
parameters which we have found to reproduce the properties of pions in very good agreement
with the data in Ref.\cite{jaus3}:
\[
m=(m_u +m_d)/2 = 260 \, MeV 
\]
\be
\b_{\pi} = 308.8 \, MeV.
\ee
For the calculations of this Section we assumed a mass difference $m_d -m_u = 4 \, MeV$.

The momentum transfer in pion beta decay is small and the form factors can be approximated
by monopole forms
\be
F_i(q^2) = \frac{F_1(0)}{1-q^2/\L^2_i} \quad , \quad i=0,1.
\ee
The explicit calculation gives $\L_1 =719 \, MeV$ (the corresponding quantity for the
charge form factor of the pion is $\L_\pi =720 \, MeV$), while
\be
F_1(0) \approx 1 - \frac{\D m^2}{(902 \, MeV)^2} = 1- 2.0 \times 10^{-5},
\ee
from which it is seen that the effect of symmetry breaking on $F_1(0)$ is of second order, in accordance
with the Ademollo-Gatto theorem \cite{ademollo}. In contrast, $F_2(0)$ is of first order in the pion
mass difference $M_+ -M_0$ and takes the value
\be
F_2(0) = - 1.44 \times 10^{-3},
\ee
which leads to the monopole approximation (2.21) for $F_0(q^2)$ with $\L_0 =1.123 \, GeV$.

The width $\G_0$ can be obtained from Eq.(2.6) by a numerical integration over $q^2$ with
the result
\be
\G_0(\pi^+ \to \pi^0 \bar{e} \nu) = 1/\t_0 \big (1-1.2 \times 10^{-5} \big ),
\ee
where $1/\t_0$ is the approximate expression given by Eqs.(1.13) and (1.14). The correction
is essentially due to the quark structure of the pion. Obviously, the effect of the symmetry
breaking, Eq.(2.25), is largely compensated by the effect of the $q^2$-dependence of the
form factors, and the sum of all structure dependent contributions to the transition 
probabilities, Eq.(2.26), is indeed very small, and can be safely neglected.
We shall continue to analyze pion beta decay in the isospin symmetry limit $m_u=m_d=m$,
with $m$ given by (2.23).

%%%%%%%%%%%%%%%%%%%% SECTION 3 %%%%%%%%%%%%%%%%%%%%%%%%%%%%%%%%
\section{The radiative corrections of $O(\a)$ from the axial vector current}
\setcounter{equation}{0}

\hspace{0.5cm} The axial vector current essentially contributes to pion beta decay in $O(\a)$
only in the two-loop processes which are represented by the vertex correction
diagrams of Fig.2 and the exchange diagrams of Fig.3. The amplitude corresponding
to the photon-exchange diagrams of Fig.2, involving the axial vector current, is
given by
\be
T^{(\g)}_2 = \frac{G_F}{\sqrt{2}} V_{u d} \frac{\a}{4\pi^3} \int d^4 k \,
\frac{A_{\m \l} \, L^{\m \l}}{(k^2+i\varepsilon)(k^2-2lk+i\varepsilon)} \,
\frac{M^2_W}{M^2_W - (q-k)^2+i\varepsilon}.
\ee
The leptonic tensor is
\bea
L_{\m \l} &=&  \ub_{\nu}(k_{\nu}) \g_\m (1-\g_5)(-\not l +\not k +m_e)\g_\l v_e(l) 
\nonumber \\
&=& -2l_\l L_\m+k_\l L_\m+k_\m L_\l-g_{\m \l}kL +i\varepsilon_{\m\l\a\b}k^\a L^\b
\eea
and the leptonic current $L_\m$ has been defined in Eq.(2.2). The hadronic tensor 
$A_{\m\l}$ contains only the axial vector part of the weak current. Current algebra
methods have been used in Ref.\cite{abers} to derive its asymptotic behavior, which
leads to the following expression for $A_{\m\l}$:
\be
A_{\m\l}=-(Q_u+Q_d) \, i  \varepsilon_{\m\l\a\b}k^\a \,
\bra \, \pi^0 (P'') |\db \g^\b u | \pi^+ (P') \, \ket \,
\frac{i}{k^2-M^2_A} + O \Big ( \frac{1}{k^2} \Big ),
\ee
where an arbitrary hadronic mass $M_A$ is introduced to avoid a spurious infrared divergence in Eq.(3.1).
The low-energy part of $A_{\m\l}$ depends on the quark structure
of the pion and is unknown.

%*********************************************************************************************
%FIGURE 2a,b,c,d

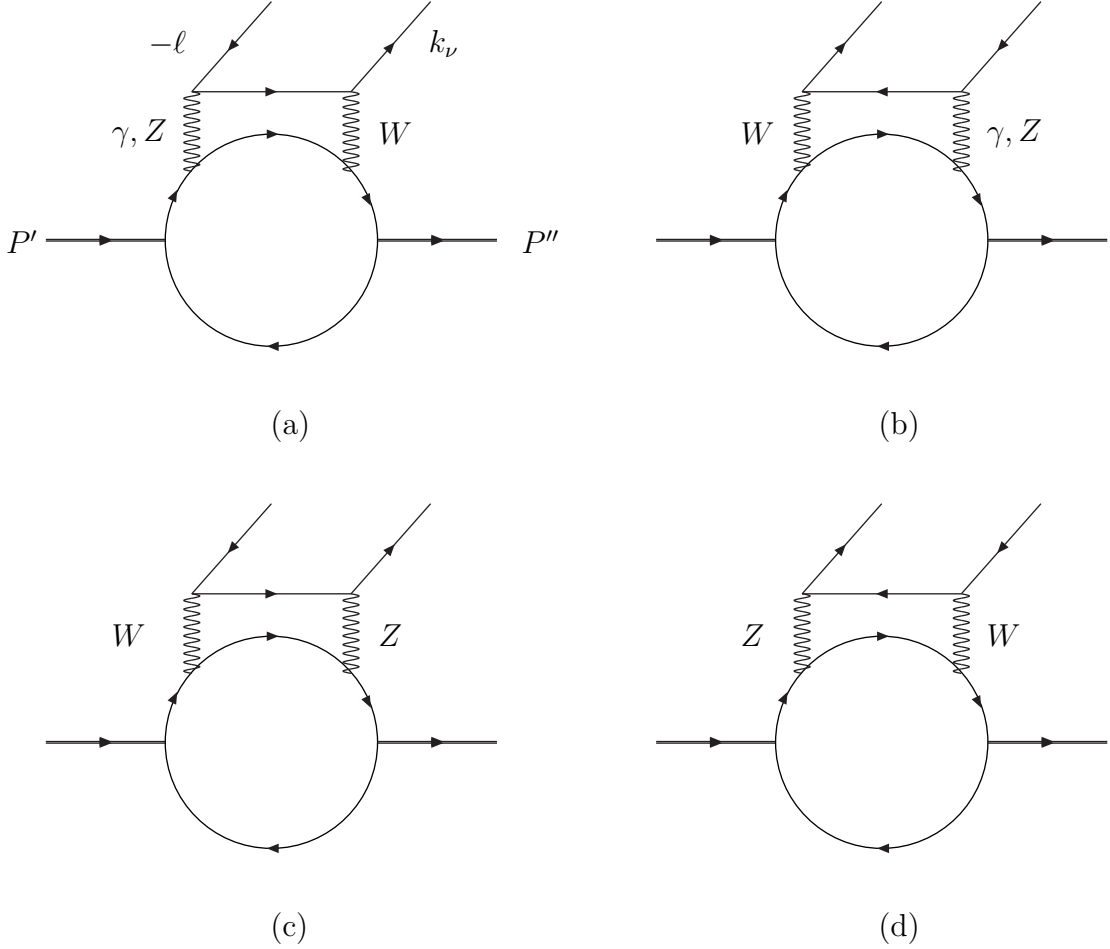
\begin{figure}[h] 
\begin{center}
\begin{picture}(400,370)

%\Line(0,0)(200,0)
%\Line(0,0)(0,180)
%\Line(0,180)(200,180)
%\Line(200,0)(200,180)

%FIGURE 2a

\put(0,190){\begin{picture}(400,180)
\Text(0,90)[l]{$P'$}
\ArrowLine(15,90.3)(60,90.3)
\ArrowLine(15,89.7)(60,89.7)
\GCirc(100,90){40}{1}
\ArrowLine(140,90.3)(185,90.3)
\ArrowLine(140,89.7)(185,89.7)
\Text(195,90)[l]{$P''$}

\ArrowLine (101,50)(100,50)
\ArrowLine (100,130)(101,130)
\ArrowLine (63,106)(64,108)
\ArrowLine (136,106)(137,104)

\Photon(70,116)(70,146){3}{9}
\Photon(130,116)(130,146){3}{9}
\ArrowLine(70,146)(130,146)

\Text(40,131)[l]{$\g,Z$}
\Text(141,131)[l]{$W$}

\ArrowLine(100,180)(70,146)
\ArrowLine(130,146)(160,180)

%\ArrowLine (115,180)(100,130)
%\ArrowLine (100,130)(165,180)

\Text(54,165)[l]{$-\ell$}
\Text(160,165)[l]{$k_\n$}

\Text(100,20)[l]{(a)}

%FIGURE 2b

\ArrowLine(245,90.3)(290,90.3)
\ArrowLine(245,89.7)(290,89.7)
\GCirc(330,90){40}{1}
\ArrowLine(370,90.3)(415,90.3)
\ArrowLine(370,89.7)(415,89.7)

\ArrowLine (331,50)(330,50)
\ArrowLine (330,130)(331,130)
\ArrowLine (293,106)(294,108)
\ArrowLine (366,106)(367,104)

\Photon(300,116)(300,146){3}{9}
\Photon(360,116)(360,146){3}{9}
\ArrowLine(360,146)(300,146)

\Text(371,131)[l]{$\g,Z$}
\Text(278,131)[l]{$W$}

\ArrowLine(300,146)(330,180)
\ArrowLine(390,180)(360,146)

\Text(330,20)[l]{(b)}
\end{picture}}

%FIGURE 2c,d

\put(0,0){\begin{picture}(400,180)

%FIGURE 2c

%\Text(0,90)[l]{$P'$}
\ArrowLine(15,90.3)(60,90.3)
\ArrowLine(15,89.7)(60,89.7)
\GCirc(100,90){40}{1}
\ArrowLine(140,90.3)(185,90.3)
\ArrowLine(140,89.7)(185,89.7)
%\Text(195,90)[l]{$P''$}

\ArrowLine (101,50)(100,50)
\ArrowLine (100,130)(101,130)
\ArrowLine (63,106)(64,108)
\ArrowLine (136,106)(137,104)

\Photon(70,116)(70,146){3}{9}
\Photon(130,116)(130,146){3}{9}
\ArrowLine(70,146)(130,146)

\Text(40,131)[l]{$W$}
\Text(141,131)[l]{$Z$}

\ArrowLine(100,180)(70,146)
\ArrowLine(130,146)(160,180)

%\ArrowLine (115,180)(100,130)
%\ArrowLine (100,130)(165,180)

\Text(100,20)[l]{(c)}

%FIGURE 2d

\ArrowLine(245,90.3)(290,90.3)
\ArrowLine(245,89.7)(290,89.7)
\GCirc(330,90){40}{1}
\ArrowLine(370,90.3)(415,90.3)
\ArrowLine(370,89.7)(415,89.7)

\ArrowLine (331,50)(330,50)
\ArrowLine (330,130)(331,130)
\ArrowLine (293,106)(294,108)
\ArrowLine (366,106)(367,104)

\Photon(300,116)(300,146){3}{9}
\Photon(360,116)(360,146){3}{9}
\ArrowLine(360,146)(300,146)

\Text(371,131)[l]{$W$}
\Text(278,131)[l]{$Z$}

\ArrowLine(300,146)(330,180)
\ArrowLine(390,180)(360,146)

\Text(330,20)[l]{(d)}
\end{picture}}

\end{picture}
\caption{Vertex corrections for pion beta decay.}
\end{center}
\end{figure}

%**********************************************************************************************

If the result (3.3) is inserted into Eq.(3.1) for $T^{(\g)}_2$ and added to
the corresponding $Z$-exchange contribution $T^{(Z)}_2$ of Fig.2, one obtains the 
correction terms of $O(\a)$ in Eq.(1.2) that are induced by the axial vector current,
where the unknown low-energy contribution is parametrized in terms of the constant $M_A$.

It is the main purpose of this work to evaluate those contributions of the vertex correction and
exchange diagrams that come from the axial vector current, in the light-front quark model of
Ref.\cite{jaus3}. The model calculation coincides with the result of Eq.(3.3) in the
asymptotic limit and completes the current algebra approach by filling in the details that
depend upon the quark structure of the pion.

In addition we shall estimate the contribution of higher order gluon exchange by means
of the $\r$ exchange diagrams of Fig.4

\subsection{The vertex corrections of Fig.2 }

\hspace{0.5cm}We shall calculate the contribution of the vertex correction diagrams of Fig.2 
by treating separately the vertex correction for an off-shell quark. In the limit
$l = k_\n =q=0$ the matrix element for the exchange of a photon, that consists of a part
that describes the $u \to d$ transitions of Figs.2a,b and an analogous part for the
$\bar{d} \to \bar{u}$ transitions, is given by
\be
T^{(\g)}_2 = G_F V_{ud}\frac{N_c}{16\pi^3} \int^1_0 dx \int d^2 p'_\perp \frac{h'^2_0}
{(1-x)N'^2_1}\frac{\a}{4\pi} \Big [Q_u \Pi^{(a)}+Q_d \Pi^{(b)} \Big ]
\ee
and the contribution of the vertex correction is 
\bea
\Pi^{(a)}& =& \frac{i}{\pi^2} \int d^4 k \,
\frac{S^{(a)}_{\m \l} \, L^{\m \l}}{(k^2+i\varepsilon)^2(k^2-2p'_1k+N'_1+i\varepsilon)} \,
\frac{M^2_W}{M^2_W - k^2+i\varepsilon}, \\
\Pi^{(b)} &=& \frac{i}{\pi^2} \int d^4 k \,
\frac{S^{(b)}_{\m \l} \, L^{\m \l}}{(k^2+i\varepsilon)^2(k^2+2p''_1k+N''_1+i\varepsilon)} \,
\frac{M^2_W}{M^2_W - k^2+i\varepsilon},
\eea 
with
\bea
S^{(a)}_{\m \l} &=& tr\left[\g_5(\not p''_1+m''_1)(-\g_\m \g_5)(\not p'_1 -\not k +m)\g_\l
(\not p'_1+m'_1)\g_5(-\not p_2+m_2)
\right], \nonumber \\
S^{(b)}_{\m \l} &=& tr\left[\g_5(\not p''_1+m''_1)\g_\l(\not p''_1 +\not k +m)(-\g_\m \g_5)
(\not p'_1+m'_1)\g_5(-\not p_2+m_2)
\right], \nonumber
\eea
where $p''_1=p'_1$ and $N''_1=N'_1$, since $q=0$. The evaluation of the traces gives the result
\bea
S^{(a)}_{\m\l}=-i\varepsilon_{\m\l\a\b} \Big (4N'_1 p'^\a_1 p^\b_2 +k^\a S^\b \Big ), \nonumber \\
S^{(b)}_{\m\l}=-i\varepsilon_{\m\l\a\b} \Big (-4N'_1 p'^\a_1 p^\b_2 +k^\a S^\b \Big ), \nonumber
\eea
where $S^\b$ has been defined in Eq.(2.14). Only the term $i\varepsilon_{\m\l\a\b}k^\a L^\b$
of the leptonic tensor $L_{\m\l}$, Eq.(3.2), contributes to the momentum integrals $\Pi^{(a)}$
and $\Pi^{(b)}$, which can be written as
\be
\Pi^{(a)} = \frac{2}{i\pi^2} \int d^4 k \,
\frac{k^2\cdot LS-kS\cdot kL+4N'_1(p'_1 k\cdot P'L-p'_1 L\cdot kP')}
{(k^2+i\varepsilon)^2(k^2-2p'_1k+N'_1+i\varepsilon)} \,
\frac{M^2_W}{M^2_W - k^2+i\varepsilon}, \\
\ee

\be
\Pi^{(b)}=\frac{2}{i\pi^2} \int d^4 k \,
\frac{k^2\cdot LS-kS\cdot kL-4N'_1(p'_1 k\cdot P'L-p'_1 L\cdot kP') }
{(k^2+i\varepsilon)^2(k^2+2p''_1k+N''_1+i\varepsilon)} \,
\frac{M^2_W}{M^2_W - k^2+i\varepsilon}, \\
\ee
where we have used that $p_2 = P'-p'_1$. The momentum integrals of Eq.(3.6) can be calculated
in terms of the usual space-time components by the standard Feynman parameter method. Using the
detailed results that have been collected in Appendix A we find
\bea
\Pi^{(a)}=\Pi^{(b)} \equiv  \Pi &=& 2 \Big \{ (3b_1 +b_2\, p'^2_1)LS-b_2 \, p'_1 S \cdot p'_1 L 
 \nonumber  \\
& & -4a_1 \, N'_1(p'^2_1 \cdot P'L - p'_1 L \cdot p'_1 P') \Big \},
\eea
where $a_1$, $b_1$ and $b_2$ are functions of $p'^2_1$ and are given by Eqs.(A3)-(A5).

The matrix element $T^{(\g)}_2$ for the exchange of a photon has been expressed in terms
of the light-front momentum integral (3.4) which is to be computed at the pole of the spectator
quark. However, it is well known (see e.g. Ref.\cite{jaus3} and references therein) that this
straightforward light-front representation of a hadronic matrix element is in general incomplete
and contains spurious contributions that violate Lorentz covariance. These difficulties are
a consequence of the fact that the effect of the associated zero-modes is not included. 
(Examples of  zero-mode contributions can be found in Appendix B, Eq.(B5), and in Ref.\cite{jaus3}.
A more general discussion in the context of light-front quantization is given in Ref.\cite{zero}.)
This is the zero-mode problem which in the present case can be circumvented by the decomposition
of the matrix element $T^{(\g)}_2$ into a covariant (physical) part, that is not associated with
a zero-mode, and a spurious part that is cancelled by the appropriate zero-mode contribution. We
are only interested in the physical part of $T^{(\g)}_2$ that can be identified by choosing
a special representation of the 4-vector $L$:
\be
L=(L^-,0,0_\perp). 
\ee
The method which we have developed in Ref.\cite{jaus3} can be used to show that the resulting
expression for $T^{(\g)}_2$ is unique, i.e. the contribution of the associated zero-mode
vanishes exactly.

We thus conclude that the condition (3.10) guarantees that all spurious contributions are
eliminated and the momentum integral (3.4), calculated at the pole of the spectator quark,
uniquely defines the complete light-front representation of the matrix element $T^{(\g)}_2$.
In order to express the quantity $\Pi$, Eq.(3.9), in terms of light-front variables, we
compute the following scalar products, using Eqs.(2.15)-(2.17), 
\bea
p'_1 L &=& x \, P'L, \nonumber \\
L S &=& 4 x M'^2_0 \, P'L, \nonumber \\
p'_1 P'&=& \frac{1}{2} (M^2_\pi + N'_1),  \\
p'_1 S &=& 2(p'^2_1 \, M^2_\pi + m^2 M^2_\pi -N'^2_1 ). \nonumber 
\eea
In this manner one finds $\Pi$ as a function of $p'^2_1=m^2 +N'_1 =m^2 +x(M^2_\pi -M'^2_0)$:
\be
\Pi = 8 P'L \Big \{ 3b_1 xM'^2_0 -(a_1 +b_2)N'_1 \Big [p'^2_1 -\frac{x}{2} (N'_1 +M^2_\pi) 
\Big ] \Big \}.
\ee
Inserting (3.12) into Eq.(3.4) we can express the matrix element $T^{(\g)}_2$ in terms
of $T_1$, Eq.(2.1), as
\bea
T^{(\g)}_2 &=& \frac{1}{2} T_1 \, \d^{(2\g)}_{axial}, \nonumber \\
\d^{(2\g)}_{axial} &=& \frac{\a}{2\pi} (Q_u +Q_d)
\frac{N_c}{16\pi^3} \int^1_0 dx \int d^2 p'_\perp \frac{h'^2_0}
{(1-x)N'^2_1} \Big \{6b_1 xM'^2_0  \nonumber \\
& & ~~~~~~~~~~~~~~~~~~~~~~~-2(a_1 +b_2)N'_1 \Big (p'^2_1-
\frac{x}{2} (N'_1+M^2_\pi ) \Big ) \Big \}.
\eea
In arriving at Eq.(3.13) we have finally established the LFQM expression for the matrix element 
$T^{(\g)}_2$.

In order to complement the above remarks regarding the zero-mode problem, we note that the
3-dimensional light-front momentum integral (3.13) with pointlike $\pi q \bar{q}$ vertices
(i.e. for $h_0 = const.$), and the covariant 4-dimensional momentum integral that
represents the photon-exchange Feynman diagrams of Fig.2 (with the same pointlike
$\pi q \bar{q}$ vertices) are equal, which is another proof that there are no zero-mode contributions.
We can reverse this argument and conclude that the transition from the covariant
Feynman perturbation theory to the LFQM proceeds in two steps: In the first step the
manifestly covariant 4-dimensional momentum integral, that corresponds to a given
Feynman diagram, is represented exactly in terms of a 3-dimensional light-front 
momentum integral. In the second step appropriate phenomenological $\pi q \bar{q}$
vertex functions are introduced into the light-front representation.

For the amplitude of the $Z$-exchange diagrams of Fig.2, that can be derived by an analogous
analysis, one finds, that the quark structure contributes to the amplitude only
to order $m^2/M^2_W$ and $m^2/M^2_Z$; this is a small effect which can be safely neglected.
Therefore, the published result, that is due originally to Sirlin \cite{sirlin1},
remains essentially unchanged and is
\bea
T^{(Z)}_2 &=& \frac{1}{2} T_1 \, \d^{(2Z)}_{axial} \nonumber \\
\d^{(2Z)}_{axial} &=& \frac{\a}{2\pi} (Q_u +Q_d) \Big ( 3 \ell n \frac{M_Z}{M_W}
+ O \Big ( \frac{m^2}{M^2_W}, \frac{m^2}{M^2_Z}  \Big ) \Big ), 
\eea
where the Minimal Standard Model relation $M_W = M_Z \, cos \Theta_W$ 
($\Theta_W$ is the weak angle) has been used (see e.g. Ref.\cite{PDG98}).

In the calculation of the vertex-loop integrals the value of the mass of the internal
quark line is important only in the low photon momentum range, where the quark mass is
essentially equal to the constituent mass. Therefore,
for the numerical computation of Eq.(3.13), we take the values (2.22) for the parameters
$m$ and $\b_\pi$, and obtain the result
\bea
\d^{(2\g)}_{axial}&=&  \frac{\a}{2\pi} (Q_u +Q_d) \Big ( 3 \ell n \frac{M_W}{m} + 
\frac{9}{4}+\D+ O \Big ( \frac{m^2}{M^2_W} \Big )  \Big ), \nonumber \\
\D &=& - 4.498,
\eea
where we have used the normalization condition (2.21).
 
The combined correction is
\be
\d^{(2)}_{axial} = \d^{(2\g)}_{axial} + \d^{(2Z)}_{axial}
 =  \frac{\a}{2\pi} \Big ( 3(Q_u +Q_d) \ell n \frac{M_Z}{m} + 
\frac{3}{4}+\frac{\D}{3} \Big ).
\ee

 \subsection{The exchange corrections of Fig.3}

\hspace{0.5cm} The exchange diagrams of Fig.3 are different from the diagrams of
Figs.1 and 2 in that both quark lines are involved in the decay process, and there
is no well defined spectator quark line.
In order to establish the LFQM expression for the
amplitude $T_3$ that corresponds to the exchange diagrams of Fig.3 we shall 
use the analysis of the covariant two-loop diagram with pointlike
$\pi q \bar{q}$ vertices ( a Feynman diagram) presented in Appendix B, as a guide.

%***************************************************************************************************
%FIGURE 3a,b

\begin{figure}[h] 
\begin{center}
\begin{picture}(400,180)

%\Line(0,0)(200,0)
%\Line(0,0)(0,180)
%\Line(0,180)(200,180)
%\Line(200,0)(200,180)

%FIGURE 3a

\Text(0,90)[l]{$P'$}
\ArrowLine(15,90.3)(60,90.3)
\ArrowLine(15,89.7)(60,89.7)
\GCirc(100,90){40}{1}
\ArrowLine(140,90.3)(185,90.3)
\ArrowLine(140,89.7)(185,89.7)
\Text(195,90)[l]{$P''$}

%UPPER QUARKLINES
\Text(52,117)[l]{$p'_1$}
\Text(143,117)[l]{$p''_1$}
\Text(43,63)[l]{$-p'_2$}
\Text(138,63)[l]{$-p''_2$}
\ArrowLine (70,116)(71,117)
\ArrowLine (130,116)(131,115)

%LOWER QUARKLINES
\ArrowLine (71,62.9)(70,64)
\ArrowLine (131,65.2)(130,64)

\ArrowLine (100,130)(115,180)
\Line (165,180)(100,130)
\ArrowLine(149,167.5)(147,166)
\ArrowLine(116,142.5)(114,141)

\Text(154,164)[l]{$-\ell$}
\Text(90,156)[l]{$k_\n$}

\Photon(101,50)(130,154){1}{9}
\Text(100,20)[l]{(a)}

%FIGURE 3b

\ArrowLine(245,90.3)(290,90.3)
\ArrowLine(245,89.7)(290,89.7)
\GCirc(330,90){40}{1}
\ArrowLine(370,90.3)(415,90.3)
\ArrowLine(370,89.7)(415,89.7)

%UPPER QUARKLINES
\ArrowLine (301,117)(300,116)
\ArrowLine (361,115)(360,116)

%LOWER QUARKLINES
\ArrowLine (300,64)(301,62.9)
\ArrowLine (360,64)(361,65.2)

\ArrowLine (330,130)(345,180)
\Line (330,130)(395,180)
\ArrowLine(379,167.5)(377,166)
\ArrowLine(346,142.5)(344,141)

\Photon(331,50)(360,154){1}{9}
\Text(330,20)[l]{(b)}

\end{picture}
\caption{Exchange corrections for pion beta decay.}
\end{center}
\end{figure}
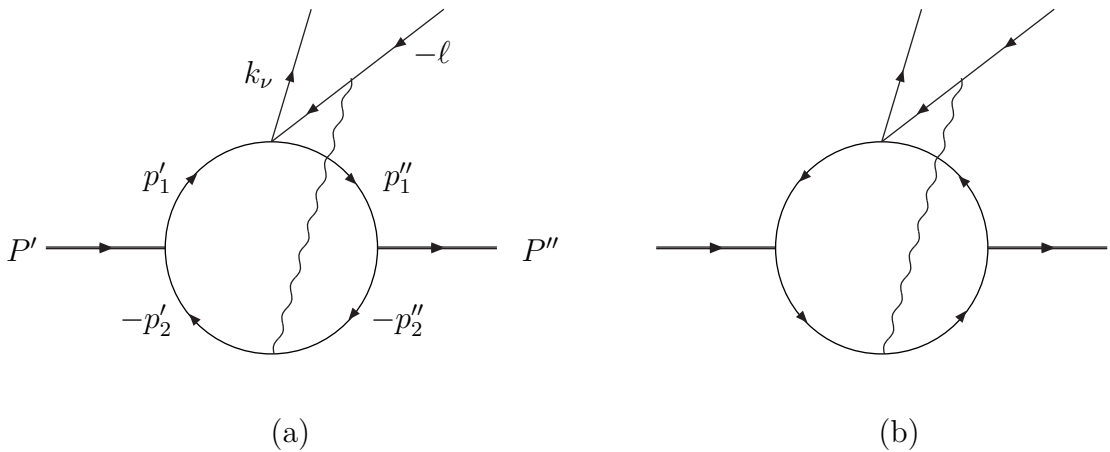

%***************************************************************************************************

Just as in the case of the one-loop momentum integrals we can calculate the two-loop
4-momentum integrals, expressed in terms of light-front variables, by performing
the integrals over the minus components of the loop momenta by contour methods,
whereby the momentum integrals are given as residues of the respective quark poles.
We have emphasized in the previous Subsection that this straightforward procedure 
leads in general to an incomplete result since it
misses the contribution of the zero-modes, and without including this effect the
contour method not only violates Lorentz covariance, but is an uncertain 
approximation of the 4-momentum integrals.
The zero-mode problem can be circumvented only if the special representation (3.10)
is used. In Appendix B we prove 
that the contribution of the zero modes vanishes exactly for the resulting amplitude $T_3$.
Therefore, the contour method already leads to the complete result for the
momentum integrals, which are given as residues of the respective quark poles,
and uniquely defines their light-front representation.

We shall calculate the amplitude $T_3$ in the limit $l=k_\n=0$, $q=l+k_\n=0$ and $P'=P''$,
with $P'^2=M^2_{\pi}$. Only the contribution of the axial vector current will be considered.
It consists of two parts, which depend upon the sign of the
plus component of the photon momentum $k=p''_1 -p'_1=p'_2 -p''_2$. 
For $k^+ > 0$ the residue is determined 
by the poles of the quarks with momenta $p'_1$ and $p''_2$. The remaining momenta are
determined by 4-momentum conservation, i.e. $p'_2 =P'-p'_1$ and $p''_1 =P'-p''_2$.  
These conditions lead to the 
relations:
\bea
\hspace{-0.6cm}
k^+ > 0 :~~~~~~ p'^2_1 - m^2 &=& 0 ~~~ ,~~~ p''^2_2 - m^2 =0 , \nonumber \\
p'^2_2 -m^2 &=& (1-x')(M^2_\pi -M'^2_0) ~~~ , ~~~  p''^2_1 - m^2 = x''(M^2_\pi -M''^2_0),
\nonumber
\eea
\be
\hspace{-1.4cm}
k^2 = k^2_{>} = \kappa \Big \{M^2_\pi -(1-x')M'^2_0 -x''M''^2_0 \Big \} -k^2_\perp,
\ee
where $\kappa = k^+ /P'^+ = x''-x'$ and $0 \le \kappa \le 1-x'$.

For $k^+ < 0$ the residue is determined by the poles of the quarks with momenta $p'_2$
and $p''_1$. These conditions lead to the relations:
\bea
\hspace{-0.6cm}
k^+ < 0 :~~~~~~ p'^2_2 - m^2 &=& 0 ~~~ ,~~~ p''^2_1 - m^2 =0 , \nonumber \\
p'^2_1 -m^2 &=& x'(M^2_\pi -M'^2_0) ~~~ , ~~~  p''^2_1 - m^2 =(1- x'')(M^2_\pi -M''^2_0),
\nonumber
\eea
\be
\hspace{-0.9cm}
k^2 = k^2_{<} = \kappa \Big \{-M^2_\pi +x'M'^2_0+(1-x'')M''^2_0 \Big \} -k^2_\perp,
\ee
and $\kappa$ varies within the range $-x' \le \kappa \le 0$. The resulting amplitude
$T_3$ is then given by
\bea
T_3 = G \, V_{ud} \frac{\a N_c}{128\pi^5}  \int^1_0 dx' \int d^2 p'_\perp
 \int^1_0 dx'' \int d^2 p''_\perp \, \frac{h'_0 \, h''_0}{x'(1-x')x''(1-x'')
(M^2_\pi -M'^2_0)(M^2_\pi -M''^2_0)} \nonumber \\
\times \Big ( \frac{\T(x''-x')}{k^4_{>}} + \frac{\T(x'-x'')}{k^4_{<}} \Big )
\bigg (Q_d S^{(a)}_{\m \l} -Q_u S^{(b)}_{\m \l} \bigg ) L^{\m \l}. \nonumber \\
\eea
The hadronic tensors associated with the diagrams of Fig.3a,b are
\bea
S^{(a)}_{\m \l} &=& tr\left[\g_5(\not p''_1+m)(-\g_\m \g_5)(\not p'_1 +m)\g_5
(-\not p'_2+m)\g_\l (-\not p''_2+m)
\right],~~~~ \nonumber \\
S^{(b)}_{\m \l} &=& tr\left[\g_5(\not p''_2+m)\g_\l(\not p'_2 +m)\g_5
(-\not p'_1+m)(-\g_\m \g_5) (-\not p''_1+m)
\right], \\
&=& -S^{(a)}_{\m \l}. \nonumber
\eea
The leptonic tensor $L_{\m \l}$ has been defined in (3.2).
The product of the hadronic and leptonic tensors consists of two parts
\be
S^{(a)}_{\m \l} L^{\m \l}= i\varepsilon ^{\m \l \a \b}k_\a L_\b S^{(a)}_{\m \l}
+ pseudoscalar,
\ee
where the pseudoscalar terms do not contribute to the amplitude $T_3$ and will be
ignored in the following presentation. The scalar terms are given by
\bea
S^{(a)}_{\m \l} L^{\m \l} &= & 8 \, (p''_1p''_2 +m^2)\bigg (p'_1k \cdot P'L-p'_1L 
\cdot P'k \bigg )
\nonumber \\
 & & -8 \, (p'_1p'_2 +m^2)\bigg (p''_1k \cdot P'L-p''_1L \cdot P'k \bigg ).
\eea
Using the special representation (3.10) we find that
\bea
S^{(a)}_{\m \l} L^{\m \l} &= & 8 \, P'L \, 
(p''_1p''_2 +m^2)\bigg (p'_1k - x' P'k \bigg )
\nonumber \\
 & & -8 \, P'L \, (p'_1p'_2 +m^2)\bigg (p''_1k - x'' P'k \bigg ).
\eea
This equation can be written such that its value at the various quark poles becomes
obvious:
\bea
S^{(a)}_{\m \l} L^{\m \l} &= & 2 \, P'L \,
\Big (M^2_\pi-(p''^2_1-m^2)-(p''^2_2-m^2)\Big )\Big ( (1-x')(p''^2_1-m^2)~~~~~~~ \nonumber \\
& & -(1-x')(p'^2_1-m^2)+x'(p''^2_2-m^2)-x'(p'^2_2-m^2)-k^2 \Big ) \nonumber \\
& & + 2 \, P'L \,
\Big (M^2_\pi-(p'^2_1-m^2)-(p'^2_2-m^2)\Big )\Big ( (1-x'')(p'^2_1-m^2) \nonumber \\
& & -(1-x'')(p''^2_1-m^2)+x''(p'^2_2-m^2)-x''(p''^2_2-m^2)-k^2 \Big ).
\eea
The contribution of $S^{(a)}_{\m \l} L^{\m \l}$ to the integrand of (3.20) for $k^+ > 0$
is, according to the conditions (3.18), given by
\be
\hspace{-5.4cm}
k^+ > 0 :~~~~~~~~~~S^{(a)}_{\m \l} L^{\m \l} = 2 \, P'L \cdot R(x',x''),~~~~~~~~~~~~~~
\ee
where
\bea
R(x',x'') &=& \Big ( (1-x'')M^2_\pi+x'' M''^2_0 \Big )
\Big ( x''(1-x')(M^2_\pi-M''^2_0)-x'(1-x')(M^2_\pi-M'^2_0)-k^2_{>} \Big ) \nonumber \\
&+& \Big ( x'M^2_\pi+(1-x') M'^2_0 \Big )
\Big ( x''(1-x')(M^2_\pi-M'^2_0)-x''(1-x'')(M^2_\pi-M''^2_0)-k^2_{>} \Big ). \nonumber \\
\eea
For $k^+ < 0$ we use the conditions (3.19) to find
\be
\hspace{-5.4cm}
k^+ < 0 :~~~~~~~~~~S^{(a)}_{\m \l} L^{\m \l} = 2 \, P'L \cdot R(1-x',1-x'').~~~~
\ee
The computation of the amplitude $T_3$ can be simplified by the observation that the
integrals for $k^+ >0$ and $k^+ <0$ are equal, which can be shown by substituting $x'$
for $1-x'$ and $x''$ for $1-x''$.

Inserting Eq.(3.26) into Eq.(3.20) we find for the LFQM expression for the matrix element $T_3$
\be
T_3 = \frac{1}{2} \, T_1 \, \d^{(3)}_{axial}
\ee
where
\bea
\d^{(3)}_{axial} &=& (Q_u +Q_d) \, \frac{\a N_c}{16\pi^5}  \int^1_0 dx' \int d^2 p'_\perp
 \int^1_{x'} dx'' \int d^2 p''_\perp \nonumber \\ 
& & \frac{h'_0 \, h''_0 \, R(x',x'')}{x'(1-x')x''(1-x'')
(M^2_\pi -M'^2_0)(M^2_\pi -M''^2_0)k^4_{>}}.
\eea
For the numerical calculation of $\d^{(3)}_{axial}$ we take again the values (2.22)
for the parameters $m$ and $\b_\pi$ and find the correction
\be
\d^{(3)}_{axial} = \frac{\a}{2\pi} \, 0.256.
\ee
   
 \subsection{The $\rho$ exchange corrections of Fig.4}

\hspace{0.5cm}Besides the diagrams of Figs.2 and 3 there is also the sum of all irreducible
higher order gluon exchange diagrams. The effect of this contribution can be
approximated by means of diagrams of the type drawn in Fig.4, where appropriate
meson states are exchanged between the weak axial vector and the electromagnetic
vertices. The Born approximation (exchange of a pion), which in general is expected 
to give the dominant contribution, vanishes, since the pion does not couple to
the axial vector current. We shall consider only the lowest mass exchange process,
i.e. the $\r$ exchange diagrams of Fig.4, and calculate the corresponding amplitude
first for on-shell vertex structures, which are defined by the appropriate
matrix elements.

%*********************************************************************************************
%FIGURE 4a,b

\begin{figure}[h] 
\begin{center}
\begin{picture}(410,180)

%\Line(0,0)(200,0)
%\Line(0,0)(0,180)
%\Line(0,180)(200,180)
%\Line(200,0)(200,180)

%FIGURE 4a
\put(-10,0){\begin{picture}(200,180)

\Text(28,80)[l]{$\pi^+$}
\ArrowLine(20,90.3)(50,90.3)
\ArrowLine(20,89.7)(50,89.7)
\ArrowLine(90,90.3)(125,90.3)
\ArrowLine(90,89.7)(125,89.7)
\ArrowLine(165,90.3)(195,90.3)
\ArrowLine(165,89.7)(195,89.7)

\Text(180,80)[l]{$\pi^0$}
\Text(105,80)[l]{$\rho^0$}

\GCirc(70,90){20}{1}

\ArrowLine (70,110)(85,160)
\ArrowLine (100,135)(70,110)
\ArrowLine (135,160)(100,135)

\GCirc(145,90){20}{1}

\Photon(145,110)(100,135){1}{9}

\Text(100,20)[l]{(a)}
\end{picture}}

%FIGURE 4b
\put(200,0){\begin{picture}(200,180)

\Text(28,80)[l]{$\pi^+$}
\ArrowLine(20,90.3)(50,90.3)
\ArrowLine(20,89.7)(50,89.7)
\ArrowLine(90,90.3)(125,90.3)
\ArrowLine(90,89.7)(125,89.7)
\ArrowLine(165,90.3)(195,90.3)
\ArrowLine(165,89.7)(195,89.7)

\Text(180,80)[l]{$\pi^0$}
\Text(105,80)[l]{$\rho^+$}

\GCirc(70,90){20}{1}

\ArrowLine (145,110)(210,160)
\ArrowLine (160,160)(152.5,135)
\ArrowLine (152.5,135)(145,110)

\GCirc(145,90){20}{1}

\Photon(70,110)(152.5,135){1}{9}

\Text(100,20)[l]{(b)}
\end{picture}}

\end{picture}
\caption{$\rho$ exchange corrections for pion beta decay.}
\end{center}
\end{figure}
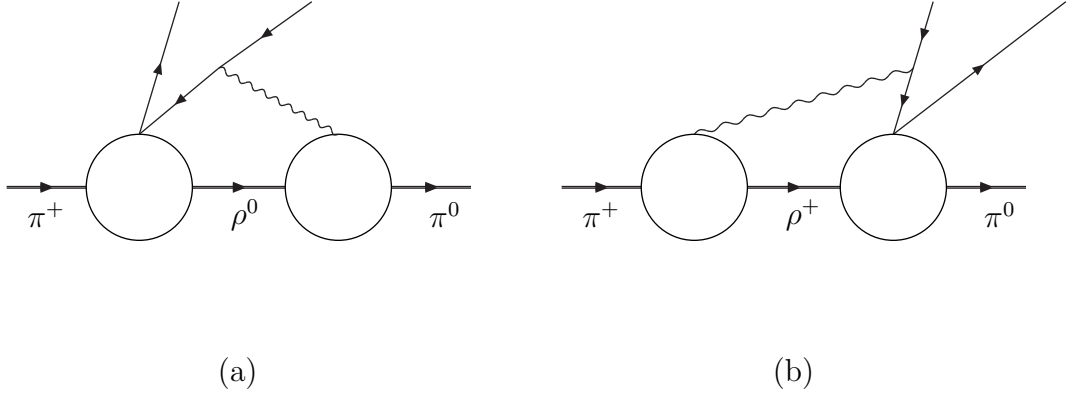

%***************************************************************************************************

The matrix element of the electromagnetic current for the transition $\r^0 \to \pi^0$ is
\bea
\lefteqn{j_\l = Q_u \ub \g_\l u +Q_d \db \g_\l d}\hspace{2.6cm} \nonumber \\
\bra P' | j_\l | P'';1J_3 \ket &=& \e^\n \G^{(\g)}_{\l \n}
\nonumber \\
&=& \e^\n \, (Q_u +Q_d)g(k^2)i\ve_{\l \n \a \b}P^\a k^\b.~~~~~~~~~~~~~~~~~
\eea
The matrix element of the axial vector current for the transition $\pi^+ \to \r^0$ is
\bea
\bra P'';1J_3 | -\db \g_\m \g_5 u | P' \ket &=& \sqrt{2} \, \e^{\ast \n} \G^{(A)}_{\l \n} \nonumber \\
&=& \sqrt{2}\, \e^{\ast \n} \, \Big \{-f(k^2)g_{\m \n}-a_+(k^2)P_\m P_\n +a_-(k^2)k_\m P_\n \Big \},
\nonumber \\
\eea
where $\e = \e(J_3)$ is the polarization vector of the $\r$, $P=P'+P''$, and
$k=P''-P'$.

The amplitude that corresponds to the diagram of Fig.4a is then given in the 
limit $l=k_\n =0$ by
\be
T_{4a} = -G \, V_{ud}(Q_u+Q_d) \frac{ie^2}{(2\pi)^4} \int d^4 k
\frac{\G^{(\g)}_{\l \n}g^{\n \r}\G^{(A)}_{\m \r}L^{\m \l}}
{(k^2+i\ve)^2((P'+k)^2-M^2_\r+i\ve)},
\ee
where we have used the bare $\r$ propagator
\bea
D_{\m \n}(P) &=& \bigg ( g_{\m \n} -\frac{P_\m P_\n}{P^2} \bigg ) \D_0 (P), \nonumber \\
\D_0^{-1} (P) &=& P^2-M^2_\r + i\ve,
\eea
and the leptonic tensor $L_{\m \l}$ is given by Eq.(3.2) and $P'^2=M^2_\pi$. Using the hadronic
tensors as defined by (3.31) and (3.32) gives the result
\be
T_{4a} =  -G \, V_{ud}(Q_u+Q_d) \frac{ie^2}{(2\pi)^4} \int d^4 k \, g(k^2)f(k^2)
\frac{4(P'k \cdot kL - k^2 \cdot P'L)}
{(k^2+i\ve)^2((P'+k)^2-M^2_\r+i\ve)}. \nonumber
\ee
In the isospin symmetry limit the amplitudes corresponding to the diagrams of
Fig.4a and 4b are equal, and the total contribution is given by
\[
T_4 = T_{4a}+T_{4b} =2 \, T_{4a}.
\]
The amplitude $T_4$ depends only on the form factors $g(k^2)$ and $f(k^2)$,
which we have determined in the framework of the light-front formalism in
Ref.\cite{jaus3}. The results of \cite{jaus3} can be written as
\bea
g(k^2;P'^2,P''^2) &=& - \frac{N_c}{8\pi^3} \int^1_0 dx \int d^2 p'_\perp 
\frac{h_\pi(M'_0) h_\r(M''_0)}
{(1-x)x^2(P'^2-M'^2_0)(P''^2-M''^2_0)} \nonumber \\
& & ~~~ \times \bigg \{m + \frac{2}{M''_0 +2m}\bigg [ p'^2_\perp + \frac{(p'_\perp k_\perp)^2}
{k^2} \bigg ] \bigg \}, \\
f(k^2;P'^2,P''^2) &=&  \frac{N_c}{8\pi^3} \int^1_0 dx \int d^2 p'_\perp 
\frac{h_\pi(M'_0) h_\r(M''_0)}
{(1-x)x^2(P'^2-M'^2_0)(P''^2-M''^2_0)} \nonumber \\
& & \times \bigg \{-2xmM'^2_0-2xmP'^2-mkP+mk^2+2m(k^2-kP)\frac{p'_\perp k_\perp}{k^2} \nonumber \\
& & -2 \frac{ p'^2_\perp + \frac{(p'_\perp k_\perp)^2}{k^2}}{M''_0 +2m}
\bigg [2xP'^2 +2xM'^2_0 -k^2 +kP-2(k^2-kP) \frac{p'_\perp k_\perp}{k^2}
\bigg ] \bigg \}, \nonumber \\
\eea
where $p''_\perp =p'_\perp -(1-x)k_\perp$, $k^2_\perp =-k^2$ and $kP = P''^2 -P'^2$. 
We have designated the $\pi q \bar{q}$ vertex function $h'_0$
by $h_\pi(M'_0)$  and the $\r q \bar{q}$  
vertex function $h''_0$ by $h_\r(M''_0)$; they are
given by Eq.(2.18) in terms of the $\pi$ mass $M_\pi$ and $\r$ mass $M_\r$, respectively.
The on-shell form factors are given by
\be
g(k^2) = g(k^2;M^2_\pi,M^2_\r)~~~,~~~~
f(k^2) = f(k^2;M^2_\pi,M^2_\r).
\ee
For the evaluation of the 4-momentum integral (3.35) it is convenient to approximate
the form factors by monopole forms:
\be
g(k^2)=\frac{g(0)}{1-k^2/\L^2_g}~~~,~~~~ 
f(k^2)=\frac{f(0)}{1-k^2/\L^2_f},
\ee
where the pole masses $\L_g$ and $\L_f$ are determined by the derivatives of the
form factors at $k^2=0$. For the numerical calculation we take the values (2.22)
for the parameters $m$ and $\b_\pi$, and $\b_\r = 0.26 \, GeV$ \cite{jaus3}, and
obtain the results
\bea
g(0)& = & -1.21 \, GeV^{-1} ~~, ~~~~ \L_g = 0.664 \, GeV , \nonumber \\
f(0) &=& -0.85 \, GeV ~~, ~~~~ \L_f = 1.72 \, GeV.
\eea
The 4-momentum integral (3.35) can now be calculated by the standard Feynman
parameter method, with the result 
\bea
T_4 &=& \frac{1}{2} T_1 \, \d^{(4)}_{axial}, \nonumber \\
\d^{(4)}_{axial} &=& \frac{\a}{2\pi}(-0.69) =
-8.0 \times 10^{-4} ~~~~~ \mbox{(on-shell)}.
\eea
Sirlin has estimated the contribution of the diagrams of Fig.4 on
the basis of vector dominance and Weinberg sum rule arguments in Ref.\cite{sirlin1}
and found the correction to the decay rate 'to be a few times $10^{-4}$', in accordance
with (3.41).

However, the intermediate $\r$ in the diagrams of Fig.4 is off-shell, and for a rigorous 
evaluation  of the corresponding amplitude the off-shell structure of the hadronic 
vertices must be accounted for. Moreover, a consistent treatment requires the
calculation of the $\r$ self-energy operator in the same $q \qb$-loop approximation
that has been used for the calculation of the hadronic form factors. The corresponding
renormalized $\r$ propagator is obtained from (3.34) by the modification
\[
\D_0^{-1} (P) \to \D^{-1}(P) = (P^2 -M^2_\r +i\ve) \, F_\r(0;M^2_\r,P^2),
\]
where $F_\r(k^2;M^2_\r,P^2)$ is the half-off-shell charge form factor of the $\r$,
with the normalization condition $F_\r(0;M^2_\r,M_\r^2)=1$. An analogous result 
has been derived for the
renormalized pion propagator in Ref.\cite{fear}. For $k^2=0$ the charge form factors
of $\pi$ and $\r$ are given in the one-loop approximation by the same analytical
expression (with appropriate $\pi$ and $\r$ parameters) which has been derived
in Ref.\cite{jaus3} to be
\bea
F_\r(0;M^2_\r,P^2) = \frac{N_c}{8\pi^3} \int^1_0 dx \int d^2 p'_\perp 
\frac{\Big ( h_\r (M'_0) \Big )^2}
{(1-x)x^2(M_\r^2-M'^2_0)(P^2-M'^2_0)} \, xM'^2_0. \nonumber \\
\eea  
In order
to estimate the importance of this structure effect we have used the light-front
formulas (3.36) and (3.37) to continue the form factors $g(k^2)$ and
$f(k^2)$ to their off-shell forms $g(k^2;M^2_\pi,(P'+k)^2)$ and $f(k^2;M^2_\pi,(P'+k)^2)$,
with $P'^2=M^2_\pi$. For an order of magnitude estimate it is sufficient to compare
the values of the different form factors at $k=0$. We find
\[
g(0;M^2_\pi,M^2_\pi)=0.125 \, GeV^{-1} ,~f(0;M^2_\pi,M^2_\pi)=-0.028 \, GeV,~
F_\r(0;M^2_\r,M_\pi^2)=-0.070,
\]
and consequently
\be
g(0;M^2_\pi,M^2_\pi)f(0;M^2_\pi,M^2_\pi)/\Big (g(0)f(0)F_\r(0;M^2_\r,M_\pi^2) \Big ) = 0.048.
\ee
Therefore, without going into any further computational details one can conclude
that the resulting correction is reduced by at least the factor (3.43), i.e.
\be
\d^{(4)}_{axial} = O \Big ( 10^{-5} \Big ) ~~~~~ \mbox{(off-shell)}.
\ee
Evidently the correction due the $\r$ exchange diagrams of Fig.4 are very small,
and can be safely neglected.

%%%%%%%%%%%%%%%%%%%% SECTION 4 %%%%%%%%%%%%%%%%%%%%%%%%%%%%%%%%
\section{Concluding remarks}
\setcounter{equation}{0}

\hspace{0.5cm}We have developed in this work a strategy to deal with
the effect of the hadronic structure on the corrections
to pion beta decay in $O(\a)$ due to the axial vector current. It is based
on the light-front quark model for the pion, whose $q\bar{q}$ 
bound state structure is well described by two adjustable parameters, constituent
quark mass and confinement scale, that have been shown in Refs.\cite{jaus2}, \cite{jaus3}
to describe a large body of data. 
In most applications of the light-front quark
model the effect of the hadronic structure could be well approximated by
one-loop diagrams. However, the axial vector $O(\a)$ corrections to the
beta decay of a $q\bar{q}$ bound state requires the calculation of 
two-loop diagrams, and we have considered three different types which
are represented by the graphs of Figs.2, 3 and 4. The
corresponding amplitudes can be expressed in terms of light-front momentum
integrals, which we have shown to be unique, i.e. there are no associated
zero-mode contributions.

The results of the quark model calculation for the corrections of  
$O(\a)$ due to the axial vector current are given by Eqs.(3.16), (3.30) and
(3.44), which can be rewritten in the form of the standard representation (1.2)
in terms of the quantities $M_A$ and $C$:
\bea
\d^{(2)}_{axial}+\d^{(3)}_{axial} &=& \frac{\a}{2\pi} \, 3(Q_u +Q_d)\ell n\frac{M_Z}{M_A},
\nonumber \\
M_A &=& 425.68 \pm 8 \, MeV,
\eea
and
\bea
\d^{(4)}_{axial} &=& \frac{\a}{2\pi} 3(Q_u +Q_d) \, 2C, \nonumber \\
C &=& O\Big ( 10^{-2} \Big ).
\eea
Comparison with (1.3) shows that the value (4.1) for the effective mass $M_A$
is clearly different from the $a_1$ meson mass, but close to the confinement scale
of the $q\bar{q}$ pion, and is even below the guessed range (1.3).
The error bar associated with $M_A$ in Eq.(4.1) is
due to the small uncertainties of the constituent quark mass $m=260\pm5$ MeV \cite{jaus2}
and the $Z$ mass $M_Z=91.188 \pm 0.007$ GeV \cite{PDG98}. The value of $C$, i.e. the
correction due to the $\r$ exchange diagrams of Fig.4 is so small that it can be neglected.
The same is true of the effect of the form factors of the pion which we have
discussed in Sect.2.

Thus our model calculation of the axial vector contribution to the radiative corrections
of $O(\a)$ not only gives a definite value for $M_A$, Eq.(4.1), which leads to larger
corrections, but also removes the
large uncertainty in the RC due to the assumed range (1.3). Using the average value
of Ref.\cite{PDG98} $\D =M_+ -M_0 =4.5936 \pm 0.0005$ MeV, which gives the end-point
energy $E_0= (M_+ +M_0)\D /2M_+ =4.5180 \pm 0.0005$ MeV, we find 
$(\a /2\pi)g(E_0)=1.0515\times 10^{-2}$ and obtain from Eqs.(1.2) and (4.1) the
value of the RC to pion beta decay
\be
\d =( 3.230 \pm 0.002) \times 10^{-2},
\ee
where the error comes from the uncertainty of $M_A$, Eq.(4.1). We emphasize that the
error on the RC is of the same order as the neglected corrections (2.27), due to the
weak form factors of the pion, and $\d^{(4)}_{axial}$, Eq.(3.44).  

We shall investigate the effect of the detailed structure of the three quark bound
state on the properties of superallowed nuclear decays and neutron decay in a future
work, in a similar manner as for the pion. If we tentatively assume that the 
effective mass $M_A$ associated with a $qqq$ nucleon will be
found approximately equal to the result given by Eq.(4.1), it is easy to show that the
unitarity sum derived from nuclear decays becomes $V^2=0.9956 \pm 0.0011$, i.e. 
the violation of unitarity seems to be much more pronounced than indicated by Eq.(1.6).

However, a definite judgement of the unitarity problem will be possible only if the
complete results of both the detailed structure calculation for $qqq$ nucleons and
the precision measurement of pion beta decay will be available.

\subsection*{Acknowledgements}
\hspace{0.5cm} I would like to thank G. Rasche and W. S. Woolcock for helpful discussions,
and A. Gashi for technical assistance.

\begin{appendix}
\section{Appendix: The momentum integrals for Sect.3.1}
\setcounter{equation}{0}

\hspace{0.5cm}  The momentum integrals encountered in the calculation of the vertex correction
diagrams of Fig.2 can be evaluated by the standard
Feynman parameter method. Convenient formulas for the on-shell case can be found e.g. in
Ref.\cite{yokoo}. In Sect.3.1 the vertex correction is calculated for off-shell quarks 
in terms of the function $\Pi$, and
the relevant integrals are given below:
\be
\frac{1}{i\pi^2} \int d^4 k \,
\frac{k_\m}{(k^2+i\varepsilon)^2(k^2\pm 2p'_1k+N'_1+i\varepsilon)} \,
\frac{M^2_W}{M^2_W - k^2+i\varepsilon} = \pm p'_{1\m} a_1, \\
\ee

\be
\frac{1}{i\pi^2} \int d^4 k \,
\frac{k_\m k_\n}{(k^2+i\varepsilon)^2(k^2\pm 2p'_1k+N'_1+i\varepsilon)} \,
\frac{M^2_W}{M^2_W - k^2+i\varepsilon} = g_{\m\n}b_1 + p'_{1\m} p'_{1\n} b_2, \\
\ee
where
\bea
a_1 &=& -\frac{1}{p'^2_1} -\frac{m^2}{p'^4_1} \ell n \frac{m^2-p'^2_1}{m^2}
+ O \Big ( \frac{1}{M^2_W} \Big ), \\
b_1 &=& \frac{1}{4} \ell n \frac{M^2_W}{m^2} +\frac{3}{8} +\frac{m^2-p'^2_1}{4p'^2_1}
+\frac{m^4 -p'^4_1}{4p'^4_1} \ell n \frac{m^2-p'^2_1}{m^2} + O \Big ( \frac{m^2}{M^2_W} \Big ), \\
b_2 &=& -\frac{1}{2p'^2_1}-\frac{m^2-p'^2_1}{p'^4_1}
-m^2\frac{m^2 -p'^2_1}{p'^6_1} \ell n \frac{m^2-p'^2_1}{m^2} + O \Big ( \frac{1}{M^2_W} \Big ).
\eea

\section{Appendix: The two-loop diagram and its zero-mode contribution}
\setcounter{equation}{0}

\hspace{0.5cm}  We shall analyze in this Appendix the two-loop diagrams of Fig.3
for the special case of pointlike $\pi q \bar{q}$ vertex functions. From the 
corresponding covariant amplitude, which is given by the Feynman rules,
we shall derive the light-front amplitude by integrating over the minus
components of the momentum variables.
For the purpose 
of this calculation we put the $\pi q \bar{q}$ coupling constant equal to 1.  
In the conventional space-time
formalism the covariant amplitude in the limit $l=k_\n =0$ is given by
\bea
T_3^{Feynman} &=&  \frac{1}{2} \, G \, V_{ud} \,  \frac{i^2 e^2  N_c }{(2\pi)^8}
\int d^4 p'_1 \int d^4 p''_1 \, 
\frac{\Big (Q_d S^{(a)}_{\m \l} -Q_u S^{(b)}_{\m \l} \Big ) L^{\m \l}}
{D'_1 D'_2 D''_1 D''_2 \,  (k^2+i\varepsilon)^2} \nonumber \\
& & ~~~~~~~~~~~~~~~~~~~~~~~~~~~~~~~~~~~ \times \frac{M^2_W}{M^2_W - k^2+i\varepsilon},
\eea
where $D'_n = p'^2_n - m^2 +i \varepsilon$,  $D''_n = p''^2_n - m^2 +i \varepsilon$ for
$n=1,2$, and $p'_1 +p'_2 = p''_1 +p''_2 = P'$ with $P'^2 = M^2_{\pi}$. The photon
has momentum $k = p''_1 -p'_1 = p'_2 - p''_2$. The leptonic tensor $ L_{\m \l}$  has been defined
in (3.2), and the hadronic tensors $ S^{(a)}_{\m \l}$ and $ S^{(b)}_{\m \l}$ in (3.20).

If the momenta are decomposed into their light-front components we have
\[
d^4 p'_1 = \frac{1}{2} P'^+ dp'^-_1 dx' d^2 p'_\perp ~~~~,~~~~
d^4 p''_1 = \frac{1}{2} P'^+ dp''^-_1 dx'' d^2 p''_\perp.
\]
We are only interested in the integration with respect to $p'^-_1$ and
$p''^-_1$, and use the same technique as in Ref.\cite{jaus3}, which is based upon
the integral representation
\be 
\frac{i}{p^2-m^2+i\e}=\int^\infty_0 d\a \,e^{i\a(p^2-m^2+i\e)}.
\ee
A similar procedure has been used in Refs.\cite{yan} to investigate the relation between
the standard covariant quantum field theory and light-front field theory.

There are three basic integrals that contribute to the amplitude (B1); these are
\bea
\bigg ( \frac{i}{2\pi} \bigg )^2  \int dp'^-_1 \, \int dp''^- _1
\frac{1}{D'_1 D'_2 D''_1 D''_2 \,  (k^2+i\varepsilon)^2} \,  \frac{M^2_W}{M^2_W - k^2+i\varepsilon} =
~~~~~~~~~~~~~~~~~~~~~~~~~~~~ 
\nonumber \\
 \frac{1}{M^2_{\pi} \, x'(1-x')x''(1-x'')
(M^2_\pi -M'^2_0)(M^2_\pi -M''^2_0)}  ~~~~~~~~~~~~~~~~~~~~~~~~~
\nonumber \\ 
\times \bigg \{ \frac{\T(x''-x')}{k^4_{>}}\frac{M^2_W}{M^2_W - k^2_{>}} 
+ \frac{\T(x'-x'')}{k^4_{<}}\frac{M^2_W}{M^2_W - k^2_{<}} \bigg \},
 ~~~~~~~~~~~~~~~~~~~~~~~~~ \\
\bigg ( \frac{i}{2\pi} \bigg )^2 \int dp'^-_1 \, \int dp''^- _1
\frac{p'^-_1}{D'_1 D'_2 D''_1 D''_2 \,  (k^2+i\varepsilon)^2} \,  \frac{M^2_W}{M^2_W - k^2+i\varepsilon} =
~~~~~~~~~~~~~~~~~~~~~~~~~~~~~~~~~~~~~~ 
\nonumber \\
\frac{1}{M^3_{\pi} \, x'(1-x')x''(1-x'')
(M^2_\pi -M'^2_0)(M^2_\pi -M''^2_0)}  ~~~~~~~~~~~~~~~~~~~~~~~~~~~~~~~~~~~~~
\nonumber \\ 
\times \bigg \{ \frac{(M^2_{\pi}-x'M'^2_0)\T(x''-x')}{k^4_{>}}\frac{M^2_W}{M^2_W - k^2_{>}} 
+ \frac{(1-x')M'^2_0\T(x'-x'')}{k^4_{<}}\frac{M^2_W}{M^2_W - k^2_{<}} \bigg \},~~~~
\eea
where $k^2_{>}$ and $k^2_{<}$ have been defined in (3.17) and (3.18), and we have used that
$P'^+ = M_{\pi}$. The result for the first two integrals (B3) and (B4) coincides with the result
obtained by the contour method, i.e. in both cases one finds the residues of the respective
quark poles, and zero-modes do not contribute.

The third integral we represent as
\bea
~~\bigg ( \frac{i}{2\pi} \bigg )^2 \int dp'^-_1 \, \int dp''^- _1
\frac{p'^-_1 \, p''^-_1 }{D'_1 D'_2 D''_1 D''_2 \,  (k^2+i\varepsilon)^2} \,  
\frac{M^2_W}{M^2_W - k^2+i\varepsilon} =
~~~~~~~~~~~~~~~~~~~~~~~~~~~~~~~~~~~~  
\nonumber \\
\frac{1}{M^4_{\pi}} \bigg \{ P_1 \T(x''-x')+P_2 \T(x'-x'')
+R_1 \d(x') \d(x'') + R_2 \d(1-x') \d(1-x'') \bigg \}.~~~~~~
\eea
The form of the residue terms is obvious
\bea
P_1 &=& \frac{(M^2_{\pi}-x'M'^2_0)(1-x'')M''^2_0}{x'(1-x')x''(1-x'')
(M^2_\pi -M'^2_0)(M^2_\pi -M''^2_0) \, k^4_{>}}, \nonumber \\
P_2 &=& \frac{(1-x')M'^2_0 (M^2_{\pi}-x''M''^2_0)}{x'(1-x')x''(1-x'')
(M^2_\pi -M'^2_0)(M^2_\pi -M''^2_0) \, k^4_{<}}.
\eea
The functions $R_n = R_n(p'_\perp,p''_\perp)$ for $n=1,2$, which determine the zero-mode
contribution, are independent of $M_\pi$, but their detailed form cannot be derived by the
method used above. For example, $R_1$ is found to be the ratio of two functions, each of which
becomes zero for $x'=x''=0$. However, there is an alternative way to determine $R_1$ and
$R_2$. It uses the limiting behavior of the integral
\be
\lim_{M_\pi \to 0} \int d^4 p'_1 \int d^4 p''_1 \, \frac{p'_1 P' \cdot p''_1 P'}
{D'_1 D'_2 D''_1 D''_2 \,  (k^2+i\varepsilon)^2} \,
\frac{M^2_W}{M^2_W - k^2+i\varepsilon} = O(M^2_\pi),
\ee
which is implied by Lorentz covariance.

If we define
\bea
P^{(0)}_1 &=& \lim_{M_\pi \to 0} P_1  = \frac{1}{(1-x')x'' k^4_{>}}, \nonumber \\
P^{(0)}_2 &=& \lim_{M_\pi \to 0} P_2  = \frac{1}{x'(1-x'') k^4_{<}},
\eea
and integrate (B5) with respect to $x'$ and $x''$, then, according to (B8) the contribution
of $O(M^{-4}_\pi)$ must vanish exactly, which gives the conditions
\bea
\int^{1}_{0} dx' \, \int^{1}_{x'} dx'' \, P^{(0)}_1 + R_1 &=& 0, \nonumber \\
\int^{1}_{0} dx' \, \int^{x'}_{0} dx'' \, P^{(0)}_2 + R_2 &=& 0.
\eea
From the Eqs.(B9) for $R_1$ and $R_2$ it can be seen that $R_1=R_2$, by
substituting $x'$ for $1-x'$ and $x''$ for $1-x''$.

This derivation of the zero-mode contribution shows clearly that a residue term which is
derived by the contour method, may contain a spurious part that is not consistent with
the requirement of Lorentz covariance. It is the zero-mode contribution that cancels this
unphysical part of the residue term. This is an example of the deep connection
between the zero-mode and the Lorentz invariance of the light-front formalism.

Next, we shall decompose the product of leptonic and hadronic tensors in the 
integrand of (B1) into products of light-front components.
We use that $ S^{(b)}_{\m \l}=- S^{(a)}_{\m \l}$,
Eq.(3.20), and the result (3.23), and find
\bea
S^{(a)}_{\m \l} L^{\m \l} = 2 \, P'L \bigg \{ &-& 2p'^-_1 p''^-_1 (k^+)^2 \nonumber \\
&+& p'^-_1 \Big [ k^+(2m''^2_\perp +x''M^2_\pi)-(1-2x')M_\pi(2p''_\perp k_\perp
+x'' M_\pi k^+) \Big ] \nonumber \\
&-& p''^-_1 \Big [ k^+(2m'^2_\perp +x'M^2_\pi)-(1-2x'')M_\pi(2p'_\perp k_\perp
+x' M_\pi k^+) \Big ] + ... \bigg \}, \nonumber \\
\eea
where we have omitted all those terms that are independent of $p'^-_1$ and
$p''^-_1$.

If the decomposition (B10) and the basic integrals (B3)-(B5) are used to perform
the integration of (B1) with respect to $p'^-_1$ and $p''^-_1$, it is obvious that
the zero-mode contribution of Eq.(B5) vanishes, since the term that contains the
product $p'^-_1 p''^-_1$ is multiplied with $k^+$. The result for the momentum
integral (B1) is given by the residues of the respective quark poles, and zero-modes
do not contribute:
\bea
T_3^{Feynman} &=& G \, V_{ud} \,P'L \, (Q_u +Q_d) \, \frac{\a N_c}{16\pi^5}  \int^1_0 dx' 
\int d^2 p'_\perp
 \int^1_{x'} dx'' \int d^2 p''_\perp \nonumber \\ 
& & ~~~~~~~~ \frac{ R(x',x'')}{x'(1-x')x''(1-x'')
(M^2_\pi -M'^2_0)(M^2_\pi -M''^2_0) \, k^4_{>}} \nonumber \\
& &~~~~ \times \frac{M^2_W}{M^2_W - k^2_{>}+i\varepsilon},
\eea
where we have used that the integrals for $k^+ >0$ and $k^+ <0$ are equal, and
the function $R(x',x'')$ has been defined in Eq.(3.26).

We have proven that the covariant
Feynman integral (B1) and the light-front integral (B11) are equal. Since
the intgrals are finite, this result can be
verified by numerical calculation.   
At this stage we depart 
from the covariant (Feynman) perturbation theory and introduce phenomenological vertex
functions, Eq.(2.18), into the two-loop light-front integrals. This step gives the light-front
quark model expressions, Eqs.(3.28) and (3.29), for the amplitude $T_3$, corresponding to 
the diagrams of Fig.3, in terms of a simple convolution of light-front vertex functions.

\end{appendix}

%%%%%%%%%%%%%%%%%%%%%% References %%%%%%%%%%%%%%%%%%%%%%%%%%%%%%%%
\def\etal{et al.}
\gdef\journal#1, #2, #3, #4 { {\sl #1~}{\bf #2}\ (#3)\ #4 }
\def\pr{\journal Phys. Rev., }
\def\prd{\journal Phys. Rev. D, }
\def\prl{\journal Phys. Rev. Lett., }
\def\jmp{\journal J. Math. Phys., }
\def\np{\journal Nucl. Phys., } 
\def\pl{\journal Phys. Lett., }

\end{document}